\documentclass[10pt,twocolumn]{article}
\usepackage[a4paper, left=1.75cm, right=1.75cm, top=2.7cm, bottom=2.7cm]{geometry}
\usepackage{natbib}
\usepackage{graphicx}
\usepackage{subcaption}
\usepackage{float}
\usepackage{afterpage}
\usepackage{algorithm}
\usepackage{authblk}
\usepackage{hyperref}
\usepackage{algorithmicx}
\usepackage{algpseudocode}
\usepackage{listings} 
\usepackage{sectsty} 
\usepackage{tcolorbox} 

\usepackage{fancyhdr}

\author[1]{James C. Young\thanks{Corresponding Author: \href{jcy204@exeter.ac.uk}{jcy204@exeter.ac.uk}}}
\author[1]{Rudy Arthur}
\author[1]{Hywel T.P. Williams}

\affil[1]{Computer Science, Innovation Centre, University of Exeter, North Park Road, Exeter EX4 4RN, UK }

\begin{document}

\title{\vspace{25px}The Language of Weather: Social Media Reactions to Weather Accounting for Climatic and Linguistic Baselines\vspace*{20px}}

\maketitle

\begin{tcolorbox}[colback=gray!15,colframe=gray!15,boxrule=0pt,arc=0pt,boxsep=0pt,left=5pt,right=5pt,top=5pt,bottom=5pt]
\begin{abstract}
This study explores how different weather conditions influence public sentiment on social media, focusing on Twitter data from the UK. By considering climate and linguistic baselines, we improve the accuracy of weather-related sentiment analysis. Our findings show that emotional responses to weather are complex, influenced by combinations of weather variables and regional language differences. The results highlight the importance of context-sensitive methods for better understanding public mood in response to weather, which can enhance impact-based forecasting and risk communication in the context of climate change.
\end{abstract}
\end{tcolorbox}

\section{Introduction} \label{section:introduction}

The effect of weather on mood and subjective well-being has long been recognised \citep{schwarz1983mood}. This applies to both transient weather conditions \citep{klimstra2011come,tsutsui2013weather, connolly2013some, feddersen2016subjective} and the overall climate \citep{rehdanz2005climate,brereton2008happiness}. Social media, particularly Twitter, has been key to understanding this effect in the past decade. Research conclusively demonstrates that social media users discuss weather and weather events \citep{sakaki2010earthquake,arthur2018social,silver2019public,minor2023adverse} and that weather affects the sentiment of these discussions \citep{baylis2018weather}. This applies to both acute weather events like storms, hurricanes, and heatwaves \citep{caragea2014mapping,spruce2020using,young2021social} as well as general trends, such as the tendency for sentiment to decrease with rising humidity \citep{hannak2012tweetin,li2014nasty}.

Understanding weather-related mental health effects is crucial in light of Climate Change. Research finds that increasing temperatures are associated with increasingly negative mental and physical health \citep{wang202043,noelke2016increasing}.
The idea of Shifting Baseline Syndrome \citep{pauly1995anecdotes}, a gradual change in what is accepted as normal \citep{soga2018shifting}, applies here. \cite{moore2019rapidly} shows that the notability of extreme temperatures is decreasing over time, with the socially accepted idea of `normal weather' adjusting on a 5-year timescale. At the same time, they also find this is not accompanied by a decrease in negative sentiment, implying no adaptation is occurring. The lack of adaptation to increasing temperatures is confirmed with other measures of mental health outcomes, like suicide rate and number of emergency department visits\citep{mullins2019temperature}. 

Climate change means not only increasing temperature but also increasingly extreme and volatile weather. Social media analysis suggests that decreasing notability as ambient conditions change without accompanying harm reduction also applies to other types of weather. For example \cite{weaver2021social} finds that the same absolute wind speed is reported as stronger or weaker depending on the typical weather conditions in the local area. \citep{zhang2022not} finds that as extreme weather events increase in frequency they decrease in notability. Thus, understanding the public response to weather via social media requires accounting for different baselines across time and space.

As well as accounting for physical and climate factors, there also remain a number of technical challenges in understanding weather and climate perception from social media data. Typically, the emotional valence of a collection of social media posts (most often tweets) is measured using sentiment analysis algorithms \citep{drus2019sentiment}. Research using such methods to understand social media related to weather and climate increasingly finds that higher accuracy requires a better understanding of the context of the discussion \citep{yao2020domain,shyrokykh2023short}. For example, words such as ``active", ``erupted" and ``fiery" have very different meanings in the context of a volcanic crisis than their more common uses \citep{hickey2024social}. The term ``global warming" is used particularly by those who deny its seriousness \citep{effrosynidis2022exploring}. Sentiment analysis should also account for the regional lexical variations in social media \citep{huang2016understanding,grieve2019mapping}. For example, \cite{grieve2019mapping} finds that the term ``pissed off" is much more common in English Twitter than Scottish, whilst the opposite is true for the word ``angry". Methods that assign more negative sentiment to one term than the other are likely biased by failing to account for geographical variation in usage.

While large language models like ChatGPT \citep{radford2019language} and BERT \cite{devlin2018bert} usually achieve high accuracy on general benchmarks, they can suffer from hidden biases \citep{bhardwaj2021investigating} and still struggle with dialectical and non-standard English \cite{fleisig2024linguistic}. Transfer learning approaches can retrain large language models for specific contexts, however, such efforts can be computationally expensive and require significant training data \cite{gao2019target}. These are also `black box' models which lack explainability \citep{arrieta2020explainable}. When informing safety-critical decisions, like sending evacuation alerts, being able to explain why such decisions are made is crucial for institutional trust and compliance \citep{kim2015confidence}.

\subsection{Objectives}

This paper seeks to address these weaknesses by properly accounting for climate and cultural baselines in social media content about the weather. The aims of this study are as follows
\begin{enumerate}
    \item Apply context-sensitive methods to understand the relationship between weather and public mood, accounting for:
    \begin{enumerate}
    \item Weather-domain specific language
    \item Regional lexical variation
    \end{enumerate}
    \item Use `white box' methods, so that results are explainable.
    \item Study the effect of combinations of weather conditions on public mood e.g. high temperature, high humidity versus high temperature, low humidity.
    \item Normalise weather response to account for local conditions, allowing the effects of absolute versus relative anomalies to be disentangled.
\end{enumerate}

The structure of the paper is as follows: Section \ref{section:methodology} (Methodology) details the data collection and filtering processes for both Twitter and weather data. Section \ref{section:results} (Results) presents the findings from the analysis. Section \ref{section:discussion} (Discussion) interprets these findings, discusses their implications, and suggests avenues for future research.

\section{Methodology}\label{section:methodology}
\subsection{Twitter Data Collection}
The tweets used for this study have been collected using the academic Twitter Application Programming Interface (API) V2, which has since been depreciated. Since we are researching human responses to weather, it was necessary to filter out automated tweets. The following filtering steps were taken: 

\begin{enumerate}
\item \textbf{Thematic Collection:} The initial dataset contains all tweets from 2021 with the term `weather' in their text (case insensitive), resulting in 8,175,136 unique tweets as identified by their tweet IDs.
\item \textbf{Geolocation:} Our focus is on tweets originating from the UK. Twitter allows users to geotag their tweets with a specific location, which returns a GeoJSON bounding box, or have their location settings enabled, which returns a GeoJSON point. Given that fewer than 1\% of tweets are geotagged using these methods \citep{twitter_developer_advanced_2024}, location inference was necessary for the majority of the tweets. Following the methodology used by \citet{arthur2018social}, which builds on \citet{schulz_multi-indicator_2013}, various indicators were cross-referenced (such as tweet text, user descriptions, and user-provided locations) against gazetteers (GADM, DBpedia and Geonames \citep{globe_global_2012, auer_dbpedia_2007, geonames_geonames_nodate}) to infer the tweet's predicted location through the most probable overlapping polygon. This process identified 1,099,124 tweets as originating from the UK. 
\item \textbf{Bot Account Removal:} In this context, bots are defined as accounts that produce automated, non-human generated tweets. A simple removal method is employed here, which involves removing tweets from accounts that individually contributed to over 1\% of all tweets in the dataset (as previously utilised by \cite{arthur2018social}). This process removed five accounts (``BodatHome\raisebox{-0.3\height}{\includegraphics[height=1em]{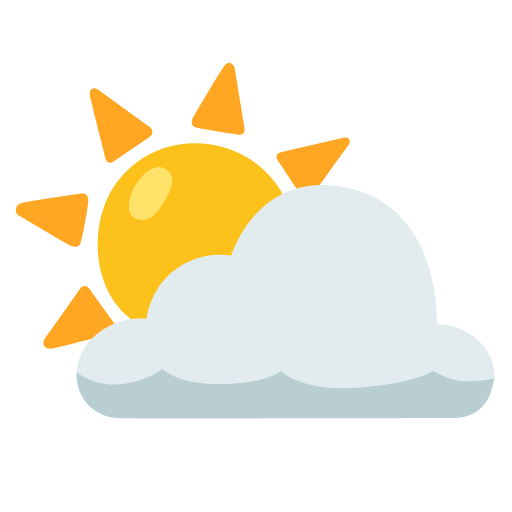}}", ``Favershamweather", ``Northampton Weather", ``Rayne Weather", ``Sigginstone Weather"), which upon manual inspection showed a high percentage of automated weather updates. After this removal, the dataset contained 1,012,319 tweets.
\item \textbf{Weather Account Removal:} Whilst the above method removed the high tweeting accounts, many smaller accounts remained in the dataset that posted high volumes of unwanted, automated tweets. These were predominantly local weather accounts providing structured updates on their regional conditions, rather than the desired, human-produced weather content. All tweets from accounts with `weather' in the username were removed, identifying 1,325 unique accounts. A manual inspection of these accounts and their tweets showed a high accuracy of this filter. This reduced the dataset to 487,151 tweets.

\item \textbf{Text Filter:} An inspection of 200 tweets showed a high percentage (18\%) of tweets that passed the above filters still contained automated weather content. These tweets were highly structured, containing a variety of weather conditions listed. They could be identified and separated through their presence of the meteorological terms mph (miles per hour) when discussing wind speed, or hPa (hectopascal) when discussing atmospheric pressure. Therefore, all tweets containing mph or hPa (case insensitive), unless found within a word (e.g., oomph or toothpaste) were eliminated. Additionally, tweets containing the phrase ``under the weather" were also removed, as this accounted for approximately 1\% of weather tweets. This final filter reduced the dataset to 401,160 tweets.

\end{enumerate}
A manual inspection of 200 random tweets showed a high relevancy of 98\% which sufficed for this investigation.

\subsection{Weather Condition Data Collection}\label{section:weatherdata}

The weather conditions investigated for this study are maximum daily values for temperature, wind speed, precipitation, humidity, and pressure. These conditions were selected because they are key indicators of weather patterns and have significant impacts on human comfort, health, and behaviour, making them crucial for understanding how weather influences public sentiment.

The weather data for this research was obtained from the E-OBS dataset, which provides a daily gridded land-only observational dataset over Europe \cite{Cornes2018}. This dataset is available on a grid with a spatial resolution of 0.25°. Although a higher resolution of 0.1° was available, the 0.25° resolution sufficed for our purposes as it was finer than the majority of the polygons obtained through location inference. After filtering to retain only grid points covering the UK, 574 daily observations were available, achieving an average daily coverage of 98.2\% across all weather conditions.

\par
Each tweet in our dataset that was successfully geolocated and filtered was then cross-referenced with the E-OBS gridded weather datasets, aligning them by time and geographical coordinates. For tweet geometries overlapping multiple grid points, the condition values were averaged; for those between points, the conditions at the nearest point were used. This process assigned every tweet in the dataset five weather conditions, estimating what the user experienced at the time and location of the posted tweet. To adjust for regional variations in weather conditions, recognising, for example, that 25°C is more common in London than in Inverness, we calculated z-scores for the weather conditions associated with each tweet. These scores normalise the data and facilitate comparisons across different regions. The z-score for each tweet's weather condition was calculated using the formula:

$$z_C(x,t) = \frac{C(x,t) - \mu_C(x)}{\sigma_C(x)}$$

In this equation, $z_C(x,t)$ represents the z-score for the weather condition $C$ at location $x$ and time $t$. $C(x,t)$ denotes the observed weather condition (temperature, pressure etc.) at the tweet's location, $\mu_C(x)$ is the average weather condition for that location over the previous 10 years (2011-2020), and $\sigma_C(x)$ is the standard deviation of the weather conditions at that location over the same period. This method ensures that the weather conditions associated with each tweet are evaluated relative to the historical weather variability of its specific location, thus normalising the data for more reliable regional comparisons.

\subsection{Linguistic Analysis}\label{sec:cider}
We used the Python library CIDER \citep{young_cider_2024} to analyse the language used within tweets. The reader is referred to the paper for the detailed discussion of this algorithm, see also \citep{hamilton2016inducing, an2018semaxis}. Briefly, this library performs domain-specific linguistic analysis by taking a text corpus and two sets of oppositely polarised seed words as inputs. CIDER then generates a custom dictionary based on the corpus which can be used to classify text. For example, to use CIDER for sentiment analysis a positive-to-negative scale is created by providing sets of positive and negative seed words e.g. \{excellent, joy, \ldots\}, \{terrible, misery, \ldots\}. The algorithm then uses a network of word associations derived from the whole corpus to discover first, second and higher relationships between the seed and other words. The output is a valence dictionary e.g. \{good:0.7, bad:-0.6, average:0.05, \ldots\}. This custom dictionary is then used, together with modifiers accounting for grammatical features of the text like negation, emphasis etc. \citep{hutto2014vader} to score sentences or, in this case, tweets.

CIDER can also be used to create scales other than sentiment, such as hot-to-cold, north-to-south, or male-to-female, by selecting seed words associated with the extremes of these dimensions e.g. the term ``ice-cream" is commonly associated with high temperatures, so would be a hot word, while ``snowman" would be on the opposite end of the hot-to-cold spectrum. The classifier built by CIDER is accurate, lightweight, and explainable and has been shown to be the best in the class of dictionary based sentiment analysis methods \citep{young_cider_2024}. While it is outperformed in terms of raw accuracy by LLMs, for understanding and synthesising a large text corpus we trade accuracy for explainability. The resulting word-level polarities can be viewed to provide context at the aggregate level and the reason for individual tweets achieving their scores can be readily determined, which would be difficult to achieve with LLMs or `black box' methods.

\subsubsection{Sentiment Analysis}
Sentiment analysis is a natural language processing (NLP) methodology that quantitatively summarises the emotion in text. For instance, ``I love the weather" may be assigned a score of +1 (positive), whilst ``This weather is horrid" may be assigned a score of -1 (negative). In this study, we analysed the sentiment of the text by training CIDER on the filtered weather tweets. The trained CIDER classifier was then used to classify individual tweets, providing sentiment scores between -1 and +1 that reflect the emotional tone of the content. The seed words selected to create this spectrum are the same as those used in the original CIDER paper \citep{young_cider_2024}:

\begin{lstlisting}[basicstyle=\fontsize{8}{10}\selectfont\ttfamily, breaklines=true, linewidth=\columnwidth, escapeinside={(*@}{@*)}]
positive_seeds = ["lovely","excellent",
    "fortunate",  "pleasant", "delightful", 
    "perfect", "loved", "love", "loves", 
    "good", "beautiful", "great", "enjoy", 
    "gorgeous",  "awesome", "nice", "amazing", 
    "excited", "(*@\raisebox{-0.3\height}{\includegraphics[height=1em]{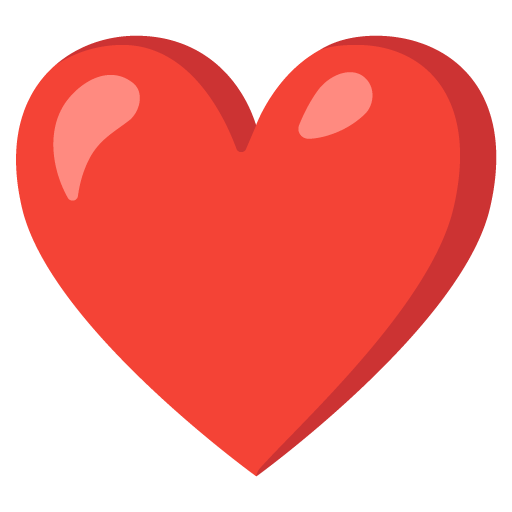}}@*)", "(*@\raisebox{-0.3\height}{\includegraphics[height=1em]{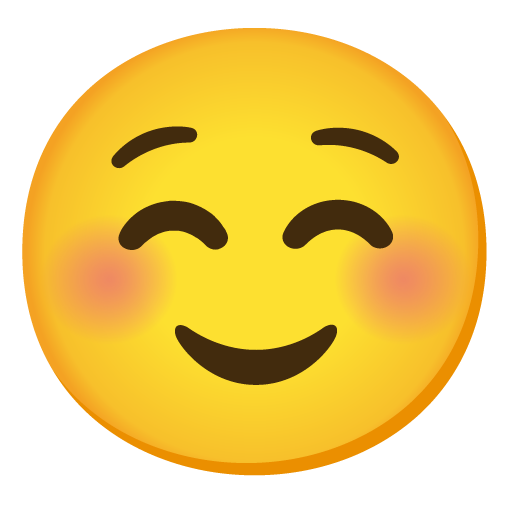}}@*)"]

negative_seeds = ["bad", "horrible", "hate", 
    "damn",  "shit", "shitty", "fuck", "hell", 
    "wtf", "hated", "stupid", "terrible", 
    "awful", "sad", "crap", "crappy", "nasty",
    "worst",  "bitch", "hates", "(*@\raisebox{-0.3\height}{\includegraphics[height=1em]{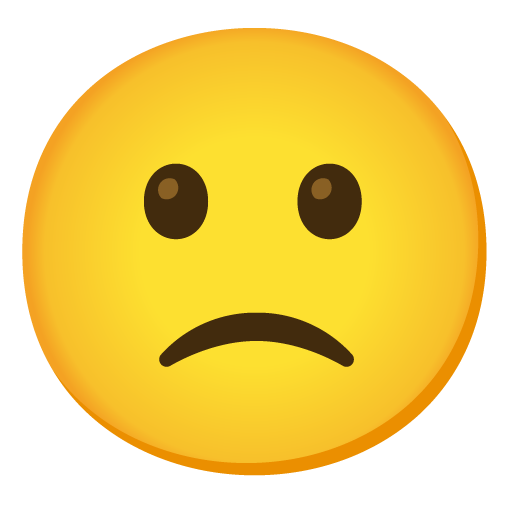}}@*)"]

\end{lstlisting}

\subsubsection{Weather Condition Lexicon Analysis}\label{sec:weatherlexicon}
CIDER was also used to quantify the valence of language on Twitter in response to different weather conditions, essentially detecting how language use changes as the weather changes. The following demonstrates how this was carried out for the maximum temperature spectrum:
\begin{enumerate}
    \item Filtered weather tweets were sorted by their associated maximum temperature z-score. 
    \item The text from tweets in the top 1\% of maximum temperature z-scores were tagged with the word ``top1", tweets in the top 1\% - 2\% were tagged with ``top2", and tweets in the top 2\% to 3\% were tagged with ``top3".
    \item Similarly, the bottom percentiles were tagged with ``low1", ``low2", and ``low3". The tags were chosen because they did not appear in the original dataset before being appended to the tweet text.
    \item CIDER was then trained on the full set of tweets (tagged and untagged), using the following seed word sets:
    \begin{lstlisting}[basicstyle=\fontsize{8}{10}\selectfont\ttfamily, breaklines=true, linewidth=\columnwidth]
    top_seeds = {`top1': 3, `top2': 2, 
                 `top3': 1}
    low_seeds = {`low1': 3, `low2': 2, 
                 `low3': 1}
    \end{lstlisting}
    Stronger weights were assigned based on the extremity of the weather condition they are associated with. This weighting scheme allowed us to generate a tailored lexical spectrum that quantifies how words are associated with the intensity of different weather conditions.
\end{enumerate}
This process was then separately repeated for precipitation, wind speed, humidity, and pressure, creating 5 distinct weather classifiers with their associated lexicons.
\section{Results}\label{section:results}

\subsection{Linguistic Variations by Weather Conditions}

\begin{figure*}[!h]
\centerline{\includegraphics[width=\textwidth]{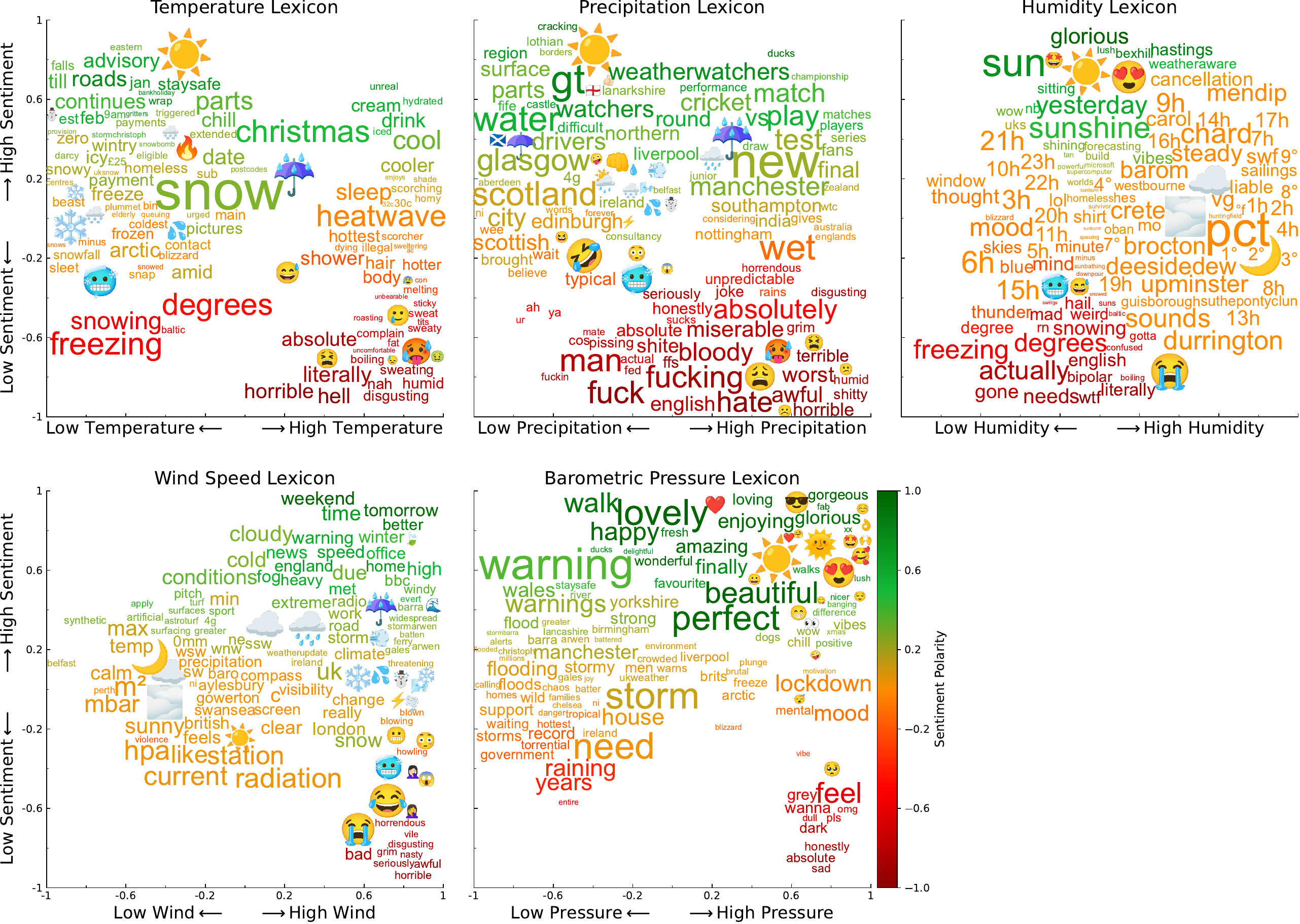}}
\caption{Linguistic variation of tweets as a result of temperature, precipitation, humidity, wind speed, and barometric pressure fluctuations, compared to sentiment fluctuations. Spectra created using CIDER \citep{young_cider_2024}.}
\label{fig:language}
\end{figure*}

As described in \ref{sec:cider}, CIDER can assign scores to words along multiple axes. In Figure \ref{fig:language} we independently trained CIDER on sentiment, maximum temperature, precipitation, humidity, pressure, and wind speed. The returned weather lexicons were then plotted against the sentiment lexicon, to see how language changes at different weather conditions and at different sentiment polarities. The y-axis shows the learned word-level sentiment and the x-axis shows the word-level scores along five weather scales. For example, the top left plot shows words scored by sentiment and temperature where the word ``freezing" indicates low sentiment (negative) and low temperature, and ``hydrated" indicates high sentiment (positive) and high temperature.

These figures show many expected associations e.g. ``sweating" associated with high temperature, ``wet" with high precipitation, positive emojis with positive sentiment, and so on, providing a sanity check of the CIDER polarities. There are also some less intuitive findings, such as the small clusters found corresponding to extreme weather conditions and very negative sentiment: low sentiment/low temperature ``baltic"; low sentiment/high wind ``grim" and low sentiment/high pressure ``sad".

\subsection{Sentiment Analysis Results}

\begin{figure}[!htbp] 
    \centering
    \includegraphics[width=\linewidth]{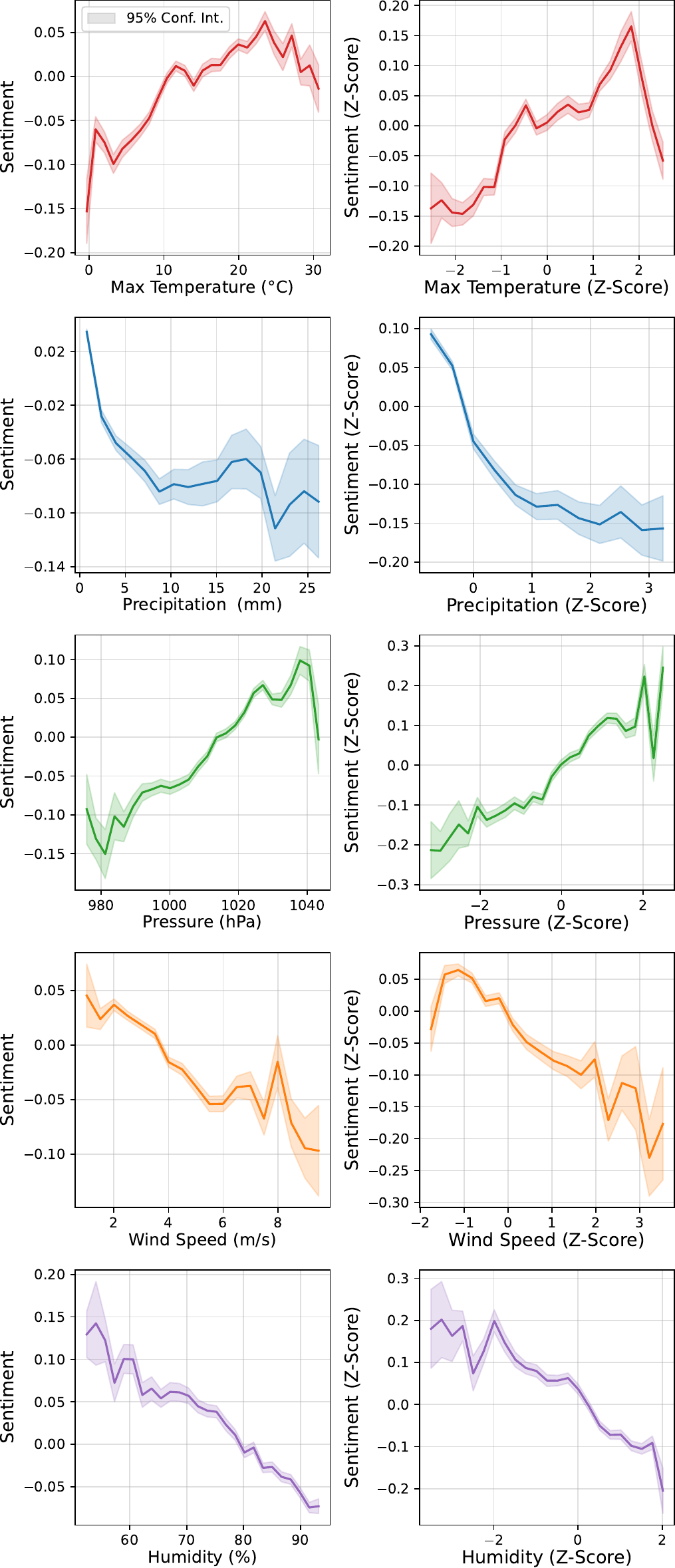}
    \caption{Change in sentiment at different weather conditions. Z-score conditions for every tweet have been calculated using historical gridded weather data \cite{Cornes2018}.}
    \label{fig:sentiment}
\end{figure}

Figure \ref{fig:sentiment} shows the relationship between sentiment and various \emph{physical} weather variables. In the left column, sentiment is plotted against raw physical measurements, e.g. the average sentiment of all tweets made at 21\textdegree C. In the right column, sentiment is plotted against the z-scores, e.g. the average sentiment of all tweets made at two standard deviations above the average temperature. Each weather condition was independently divided into 30 evenly spaced bins, and only bins containing at least 0.1\% of all tweets were plotted.
\newpage
The z-transformation is not dramatic but has some notable effects. In general, it seems to smooth the curves and emphasises trends suggested in the absolute plots. The most significant effect is on the temperature plot where sentiment rapidly increases once temperatures climb above $z \simeq 1$, peaking at $z \simeq 1.5$ and then decaying rapidly, going negative for extremely high temperatures $z > 2$. $z=2$ is a quite different absolute temperature; for instance, $z=2$ in Glasgow is 22.37$^{\circ}$C compared to 27.92$^{\circ} $C in London. This effect is not seen clearly in the left-hand `absolute' plots. 

The other weather conditions show different effects. The sentiment versus precipitation curve is much smoother when plotted against z-transformed precipitation data. Sentiment falls rapidly with any precipitation $z > 0$, saturating at around $z > 1$ and subsequently decreasing more slowly. Using the above comparison, regarding precipitation, $z=2$ equates to 15.77mm and 9.4mm in Glasgow and London respectively. The effects of the z-transformation on the shape of the pressure and wind speed response are minor. The transformation applied to humidity emphasises a sharp decrease in humidity on days that are very humid relative to the baseline. Note also that the z-transformation increases the range of sentiment scores significantly for all variables apart from temperature e.g. sentiment versus absolute humidity varies between around -0.05 and 0.2, while sentiment versus z-transformed humidity varies between around -0.2 and 0.3. This happens because using z-scores averages together tweets of similar sentiment, and tweets responding to unusual local conditions are not `diluted' by tweets from places where those conditions are normal.

The identified weather conditions do not occur independently of each other, for instance, high humidity is less common during periods of extreme wind, and high pressure often co-occurs with high temperatures. Combinations of conditions like high winds at the same time as high precipitation, i.e. a storm, are likely to elicit a unique response. Therefore, we next investigate how the sentiment of tweets changes in response to combinations of these conditions. To enable a visual inspection of the variation, a pairwise analysis is carried out, where a hexbin heatmap is plotted with the colour dictating the average sentiment of the tweets at that pair of weather conditions in the UK.

\begin{figure*}[!h]
\centerline{\includegraphics[width=0.9\textwidth]{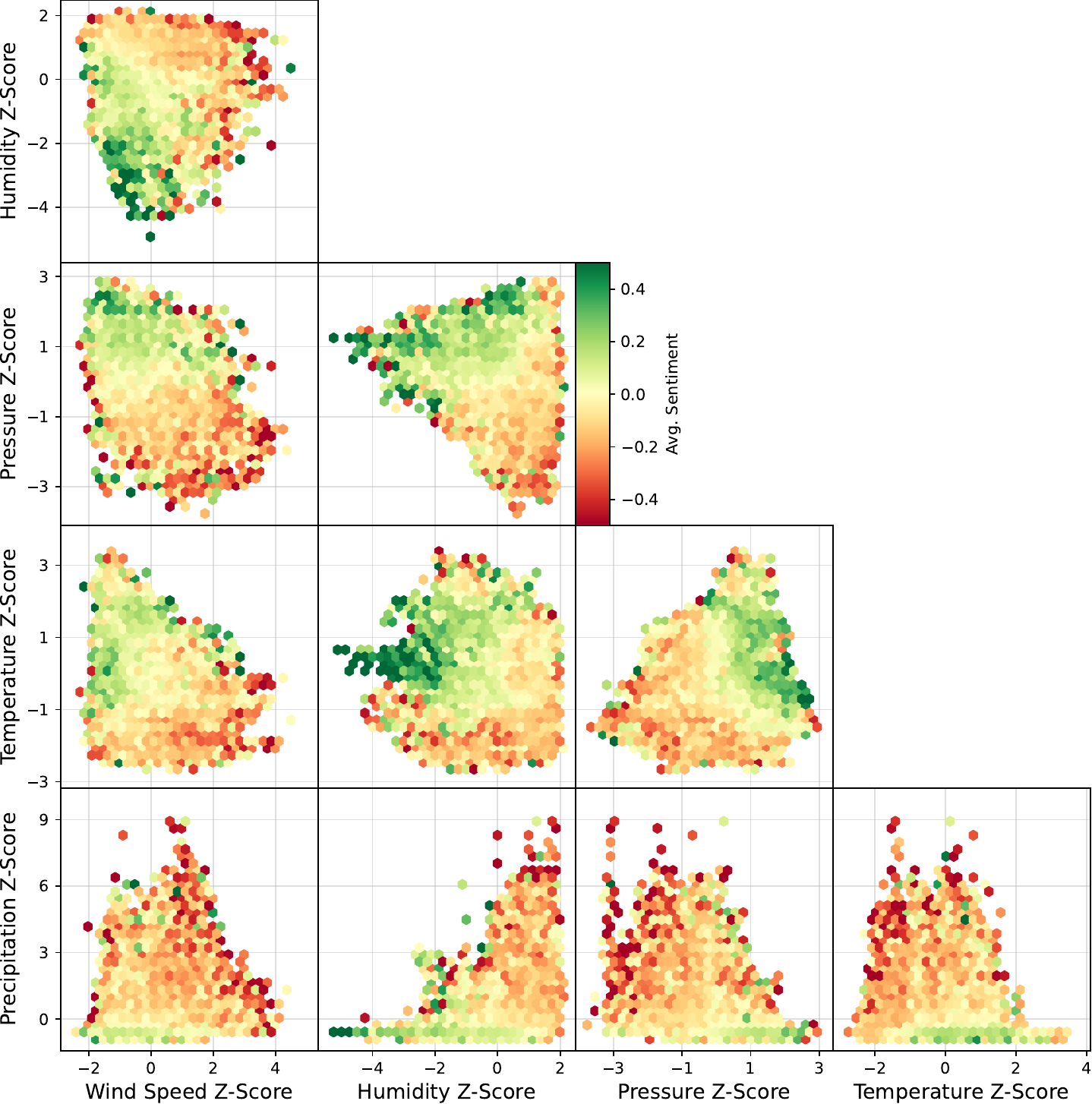}}
\caption{Hexbin plot comparing the average sentiment of tweets at different combinations of weather conditions.}
\label{fig:hexbin}
\end{figure*}
Figure \ref{fig:hexbin} shows the varying sentiment across combinations of weather conditions from tweets in the UK. Where the bin has fewer than 5 tweets, it has been omitted due to lack of volume. This is due to not all combinations of weather occurring in the data, for example, there are no days that are both high temperature (z-score > 2) and high wind (z-score > 3). For conditions that do co-occur, the response of sentiment is non-trivial. For example, at the locally average temperature, $z=0$, sentiment and humidity (third row, second column) are in an inverse relationship - high sentiment at low humidity and vice versa. At low temperatures $z < -1$, sentiment is negative for any value of humidity. On the other hand, at moderately high temperature $1<z<2$ sentiment is high regardless of humidity. At the highest observed temperatures $z>2$, there isn't a great range of humidity observed, but sentiment is negative regardless. In general, the preferred weather of UK Twitter is low-humidity, moderately high temperature, low to no wind and high pressure. The worst weather is low temperature and pressure with high wind, precipitation and humidity i.e. storms.

Together Figures \ref{fig:sentiment} and \ref{fig:hexbin} demonstrate that emotional response to weather is not straightforward - it is non-linear, multivariate, and depends on average local conditions. 

\subsection{Regional Sentiment Variations}

\begin{figure*}[!htbp]
\centerline{\includegraphics[width=0.7\textwidth]{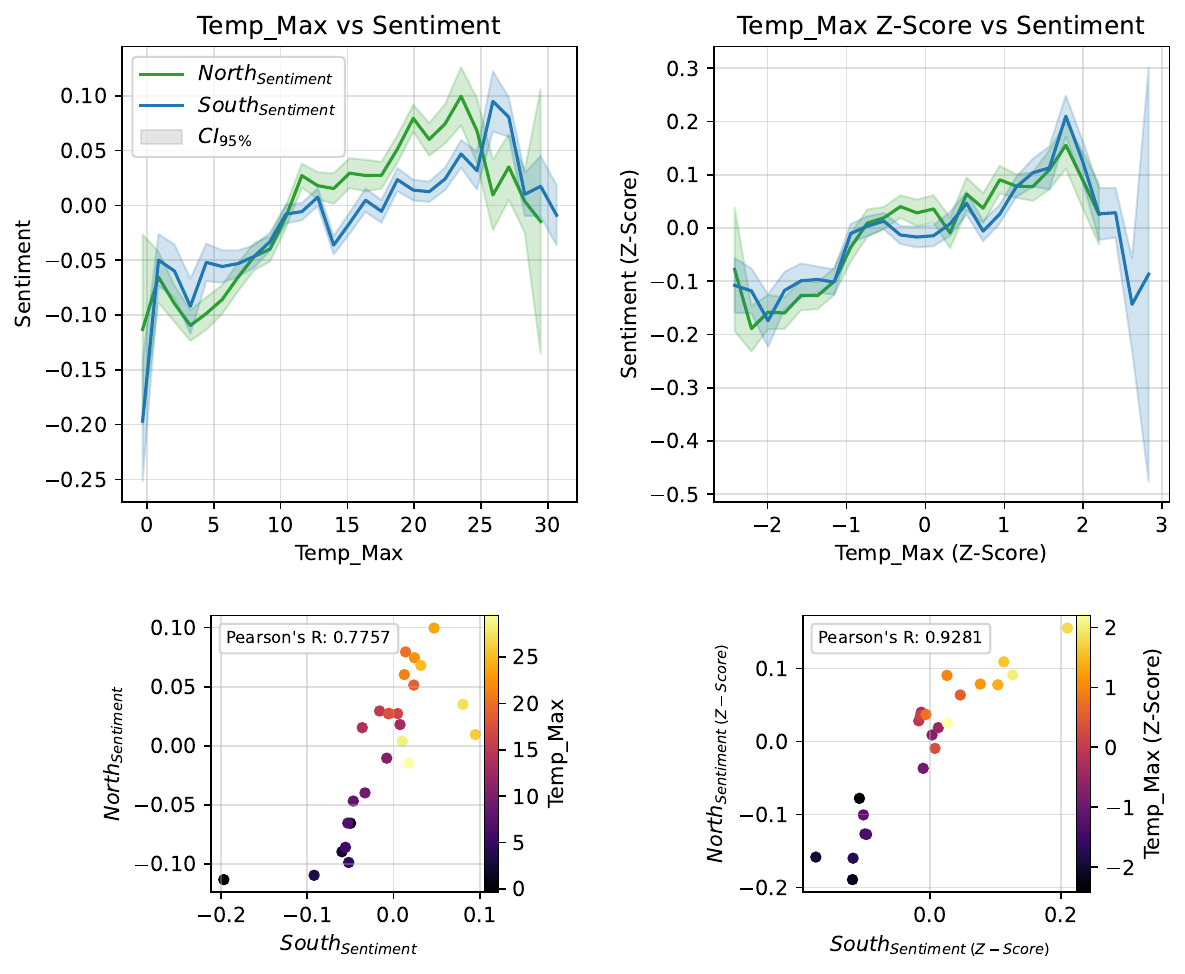}}
\caption{North vs South sentiment responses to temperature in the UK. Conditions and Sentiment have been normalised by z-scores on the right-hand side.}
\label{fig:regional_sentiment}
\end{figure*}

Next, an investigation into the regional variation of weather perception was carried out. For this, tweets were split into datasets for the North and South of the United Kingdom. To determine whether a tweet was from the North or South, tweets were assigned to a NUTS1 geographical region. If at least 50\% of a tweet's geolocated polygon (obtained from Section \ref{section:methodology}) was found within a single NUTS1 region, it was assigned to the corresponding NUTS1 region. This filtering method removed tweets that were broadly identified (e.g., "England" or "UK") and geographically ambiguous tweets. Tweets from the following regions: "Scotland," "North East England," "North West England," and "Yorkshire and The Humber" were assigned to the North, while tweets from "London," "South East England," "South West England," and "East of England" were assigned to the South. Due to the significant north-south length of the UK, there are notable differences in average winter and summer daily maximum temperatures between these regions. For instance, the average summer daily maximum temperature in the North is 18.71$^{\circ}$C, compared to 20.88$^{\circ}$C in the South. In winter, the average daily maximum temperature is 5.65$^{\circ}$C in the North and 8.56$^{\circ}$C in the South, as calculated from Section \ref{section:weatherdata} data. The North/South axis also represents the primary cultural and linguistic divide in the UK \cite{grieve2019mapping}.
\newpage
Focusing on maximum daily temperature, Figure \ref{fig:regional_sentiment} presents the same information as Figure \ref{fig:sentiment}, but with separate plots for the North and South of the United Kingdom. Similar to Figure \ref{fig:sentiment}, data points were only plotted if the bin contained at least 0.1\% of all tweets. The absolute plot (top left of Figure \ref{fig:regional_sentiment}) shows that for the same absolute temperature, the sentiment in the North is more intense (higher or lower) than the response in the South, and the positive response to warm weather peaks earlier in the North. 

Plotting the z-scores (top right of Figure \ref{fig:regional_sentiment}) brings the two curves into close agreement. Note that the agreement is worse if we only normalise temperature but not sentiment and vice versa. When accounting for local temperature baselines (with the South being warmer) and local sentiment baselines (with the North expressing more intense sentiments), we find that both regions exhibit the same response to temperature. This response includes negative sentiment at low temperatures, approximately linearly increasing up to a peak at around $z=1.5$, then a fairly rapid decline at higher temperatures. This suggests that the optimal temperature in the UK is 1.5 standard deviations above the local average, regardless of what that average is.

The bottom two plots are a statistical check of the visual comparison. Plotting the green curve against the blue one will give a straight line if the two are identical. There is a significant increase in correlation, from 0.776 ($P<0.001$) to 0.928 ($P<0.001$). The increase in correlation between the North and South expressed sentiment when the z-score is taken is also found for the other weather variables and is shown in Appendix \ref{appendix:a}.

\newpage

\section{Discussion}\label{section:discussion}
Our analysis confirms a consistent response in social media sentiment to weather, aligning with previous findings. However, our work extends this understanding by examining how baseline meteorological conditions and regional differences in climate and dialect influence this response.

Firstly, we demonstrate that a `social Beaufort scale' \citep{weaver2021social} can be constructed using CIDER and a simple algorithm for selecting seed words. Figure \ref{fig:language} illustrates these results, which also shows word-level sentiment. This approach could be applied in reverse, to infer weather conditions from tweet text. This is becoming increasingly important given the observations of \cite{moore2019rapidly} and others on the decreasing notability of extreme weather. Even if the volume of social media is no longer a useful barometer of weather impact, analysis of the text can still be used to infer both public mood and the weather conditions themselves.

We find that trends in sentiment are more clearly emphasised when local weather baselines are taken into account, as shown in Figure \ref{fig:sentiment}. This is particularly important for temperature in the UK, which spans approximately 10 degrees of latitude, resulting in significantly colder conditions in the north compared to the south. Without considering these local baselines, the response of public mood to extreme temperatures remains obscured. When we account for local baselines, we observe a significant decrease in sentiment as temperatures rise $\sim 2 \sigma$ above normal. There may be a temporal as well as a spatial component, as people's baseline expectations change over time with a warming climate. Although we currently lack a sufficiently long time series to explore this fully, future studies understanding changing responses over time are crucial for social media analysis to effectively contribute to understanding public perception and response to weather in a changing climate.

We also find that public mood is influenced in a non-linear way by combinations of weather conditions, as shown in Figure \ref{fig:hexbin}. For example, a relatively high temperature ($\sim z=1$) can be associated with both positive or negative sentiment depending on humidity levels. Climate change affects all aspects of the climate, so understanding or predicting changes in public mood requires multi-dimensional analyses. In general, weather conditions that cause particularly strong negative or positive reactions are not surprising e.g. high precipitation and high wind causing negative sentiment. However, there are intriguing, less obvious effects, such as unusually high pressure being associated with negative sentiment. This is also evident in the high-pressure cluster in Figure \ref{fig:language}, where words like ``grey, dark, vibe, feel, sad, \ldots" are prominent.

Finally, our most significant finding, illustrated in Figure \ref{fig:regional_sentiment}, demonstrates the importance of both cultural and weather baselines. The north-south axis divides the UK into distinct cultural and climate regions. The north is colder and expresses sentiment more strongly on social media than the south \citep{arthur2019human}. This results in the ideal temperature peak occurring at lower temperatures in the north than in the south, suggesting a lower temperature baseline. However, the sentiment at colder temperatures is \emph{more} negative in the north. When we account for local conditions and variations in sentiment intensity, we find that the two curves align, suggesting a universal response to weather once both baselines are considered.

This has significant implications for all analyses of social media responses to weather. Most countries and regions have a diversity of climates and languages, meaning that average expectations may not accurately reflect the experiences of any specific group. Methods like CIDER, which are flexible enough to account for local and contextual variations in communication, are crucial for this type of analysis to be useful to forecasters. For example, the move towards impact-based forecasting \cite{taylor2018communicating} emphasises the social dimensions of weather, with social media analysis having been used to validate these predictions \cite{arthur2018social,wyatt2024implementing}. Failing to account for variations in local dialects risks under- or overestimating weather impacts. Many academics and forecasting professionals aim to evaluate and improve risk communication strategies \citep{lazrus2016know}. Understanding local dialects can help not only in evaluating the social media response to such messaging but also in creating bespoke messages for different communities.

Approaches to studying weather on social media using volume and general-purpose NLP methods have generated numerous insights. Our study demonstrates that accounting for weather and linguistic baselines is crucial for achieving a more accurate and nuanced analysis. We also recognise that in a changing climate and an ever-evolving social media landscape, flexible and transparent methods are essential to ensure that the information available to academics, forecasters, and weather professionals remains relevant and useful. We believe the methods and results presented in this paper represent a significant advancement beyond previous approaches to social media analysis of weather.

\bibliographystyle{agsm}
\bibliography{references.bib}

\newpage
\appendix
\section{Extended Regional Sentiment Variations}\label{appendix:a}

\begin{figure}[htpb]
\centering
\includegraphics[width=\textwidth]{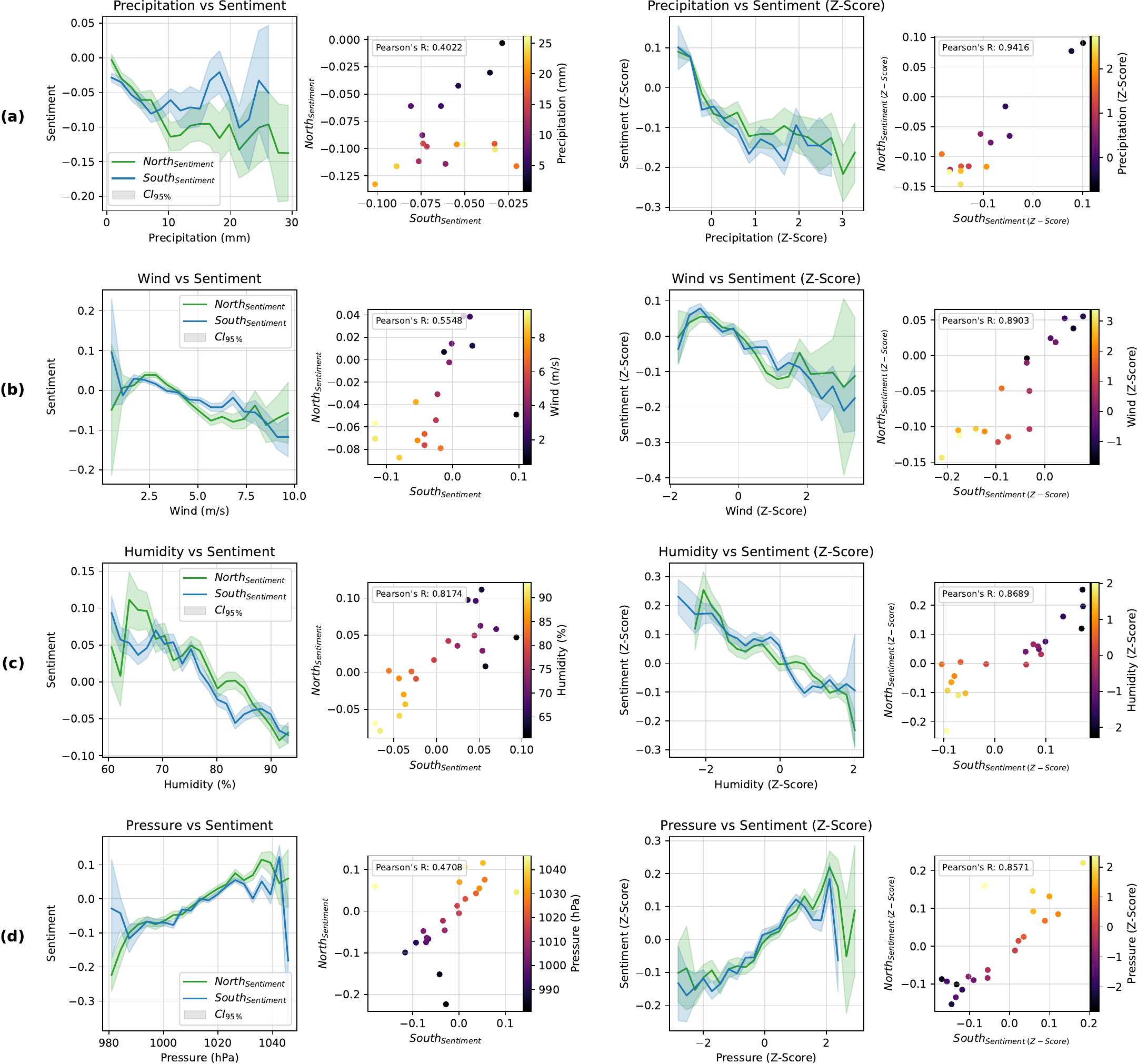}
\caption{North vs South sentiment responses to precipitation, wind, humidity, and pressure variations in the UK. Conditions and sentiment have been normalised by z-scores on the right-hand side.}
\label{fig:NS_Conditions}
\end{figure}

\end{document}